\documentclass[aps,twocolumn,groupedaddress,amsmath,amssymb]{revtex4-2}
\usepackage{graphicx}
\usepackage{color}

\begin{document}

\title{Lifshitz black holes in four-dimensional Critical Gravity.}

\author{Mois\'es Bravo-Gaete}
\email{mbravo-at-ucm.cl} \affiliation{Facultad de Ciencias
B\'asicas, Universidad Cat\'olica del Maule, Casilla 617, Talca,
Chile.}

\author{Mar\'ia Montserrat Ju\'arez-Aubry}
\email{mjuarez-at-astate.edu} \affiliation{Arkansas State University Campus Queretaro,
 Carretera estatal $\#100$ km. 17.5, Municipio Col\'on, 76270, {Quer\'etaro}, M\'exico}

\author{Gerardo Vel\'azquez Rodr\'iguez}
\email{gvelazquez-at-astate.edu} \affiliation{Arkansas State University Campus Queretaro,
 Carretera estatal $\#100$ km. 17.5, Municipio Col\'on, 76270, {Quer\'etaro}, M\'exico}

\begin{abstract}
In this work, we study the existence of asymptotically Lifshitz black holes in Critical Gravity in four dimensions {with a negative cosmological} constant under two scenarios: First, including dilatonic fields as the matter source,  \textcolor{black}{where we find} an asymptotically Lifshitz solution for \textcolor{black}{a fixed value of the dynamical exponent} $z=4$. \textcolor{black}{As a second case}, we also added a non-minimally coupled scalar field $\Phi$ with a potential given by a mass term and a quartic term. Using this approach, we found a solution for $z$ defined in the interval $(1,4)$, recovering the \textcolor{black}{Schwarzchild-Anti-de Sitter} case with planar base manifold in the isotropic limit. Moreover, when we analyzed the limiting case $z=4$, we found that
{there exists an additional solution that can be interpreted as a stealth configuration in which the stealth field is overflying the $z=4$ solution without the non-minimally coupled field $\Phi$}.
Finally, we studied the {non-trivial} thermodynamics of these new anisotropic solutions and found that they all satisfy the First Law of Thermodynamics as well as the Smarr relation. { We were also able to determine that the non-stealth configuration is thermodynamically preferred in this case.}
\end{abstract}

\maketitle
\newpage

\section{Introduction}

The idea of modifying the Einstein-Hilbert action of gravity is not new. Three-dimensional examples of this are the introduction of Topological Massive Gravity \cite{Deser:1981wh}, \cite{Deser:1982vy}, which includes a massless graviton and a massive propagating spin-2 field, and New Massive Gravity \cite{Bergshoeff:2009hq} which reproduces a unitary Fierz-Pauli theory for free massive spin-2 gravitons, at the linearized level. A few years ago, Critical Gravity, a four dimensional, ghost-free, renormalisable theory of gravity, was introduced by Lu and Pope in \cite{Lu:2011zk}. The Critical Gravity action
\begin{subequations}\label{eq:Squad}
\begin{eqnarray}
S_{g}&=&\frac{1}{2\kappa}\int{d}^4x\sqrt{-g}
\left(R-2\Lambda+\beta_1{R}^2
+\beta_2{R}_{\alpha\beta}{R}^{\alpha\beta} \right),\nonumber\\
&=&\int d^{4}x \sqrt{-g} \mathcal{L}_{\mathrm{grav}}, 
\end{eqnarray}
where $\Lambda$ is the cosmological constant, allows for the massive spin-0 field to vanish if the coupling constants $\beta_1$ and $\beta_2$ are restricted to obey the relations
\begin{eqnarray}\label{relations}
\beta_{2}=-3\,\beta_{1},\qquad \beta_{1}=-\frac{1}{2\Lambda}.
\end{eqnarray}
\end{subequations}

 Critical Gravity can naturally support the \textcolor{black}{Anti-de Sitter (AdS)} spacetime, as well as a Schwarzchild-AdS black hole configuration, {allowing us to obtain conserved quantities via the Noether-Wald method \cite{Anastasiou:2017rjf} as well as higher order extensions (as shown  in \cite{Anastasiou:2021tlv} for the six dimensional case)}. However, it was shown in \cite{Lu:2011zk} that this solution is massless which, as the authors mention \emph{might be the price to pay} to have such a well behaved theory. With all the above, we are tempted to ask the following questions: 1. Can Critical Gravity support more general spacetimes?, 2. Could it be possible, using these new asymptotic spacetimes, to find black hole configurations in Critical Gravity such that they include the Schwarzchild-AdS with planar base manifold case as a limit?. {3. If we can find such configurations, what \textcolor{black}{will happen} with their thermodynamics? Will they showcase \textcolor{black}{non-vanishing} thermodynamical quantities?}. \textcolor{black}{To answer these questions}, we will focus on the Lifshitz spacetime \cite{Kachru:2008yh}.
%
\textcolor{black}{The first motivation for choosing this particular spacetime arises from quantum phase transitions in condensed matter physics}. These transitions occur between two different phases at zero
temperature. At this critical point, the system becomes invariant
under a rescaling symmetry with eventually different weights between
the space and the time,
\begin{equation}\label{eq:symmetry_lifshitz}
t\to\tilde{\lambda}^z\,t, \qquad \vec{x}\to\tilde{\lambda}
\vec{x},
\end{equation}
where the constant $z$ is called the dynamical exponent and $\tilde{\lambda}$ is a constant.
Together with the above, the quantum critical point has \textcolor{black}{proven to be} a useful tool to understand the entire phase diagram, as well as to obtain transport coefficients (see for example \cite{Pang:2009wa,Ross:2009ar,Kuang:2017rpx,Brito:2019ose,Eling:2014saa,Lemos:2011gy,Ge:2014eba}). In general, \textcolor{black}{these quantities show to be difficult to calculate since the systems in question are usually strongly coupled systems.}
Nevertheless, the Anti-de
Sitter-Conformal Field Theory (AdS/CFT) \cite{Maldacena:1997re} correspondence (case $z=1$) has \textcolor{black}{ proved} to be a powerful tool for
studying strongly coupled systems by mapping them into classical
theories of gravity and establishing a dictionary between both
theories. However, other values of $z$ are present experimentally in
particular in the condensed matter physics. This is one the main
reasons of the recent interest of extending the AdS/CFT
correspondence to the non-relativistic physics and to condensed
matter physics. In this perspective, the Lifshitz
group, which is characterized by an anisotropic scaling symmetry but
without Galilean boosts, plays an important role. In this case, the gravity dual metric of
these systems is \textcolor{black}{referred to} as Lifshitz metric \cite{Kachru:2008yh} and
is given, in four dimensions, by
  \begin{equation}
\label{eq:Lifshitz} ds_L^2=-\frac{r^{2z}}{l^{2z}} \,dt^{2}+\frac{l^{2}}{r^{2}}\,dr^{2}+\frac{r^{2}}{l^{2}}\,\sum_{i=1}^{2}dx_{i}^{2},
\end{equation}
whose scaling symmetry is (\ref{eq:symmetry_lifshitz}) together with $r\to\tilde{\lambda}^{-1} r$.

{Unfortunately, Critical Gravity cannot support the Lifshitz spacetime (\ref{eq:Lifshitz}), let alone, asymptotically Lifshitz black holes as shown in \textcolor{black}{three dimensions \cite{Ayon-Beato:2009rgu} as well as higher dimensions} \cite{AyonBeato:2010tm}, where the authors use dimensional continuation to find the 4-dimensional branches in which asymptotically Lifshitz black holes can exist in \textcolor{black}{General Relativity} complimented with the most general quadratic corrections in the curvature. On another note, the work of Taylor \cite{Taylor:2008tg} shows that the introduction of dilatonic fields can enrich the equations of motion of General Relativity to support Lifshitz black holes, even when it is a theory whose vacuum does not support the Lifshitz spacetime in the first place. This idea is furthered  by Tarrio and Vandoren in \cite{Tarrio:2011de} \textcolor{black}{where the addition of  abelian $U(1)$ fields and one real scalar field allows to obtain analytic asymptotically \textcolor{black}{Lifshitz black hole configurations} in arbitrary dimensions with $z \geq 1$}. Inspired by these previous works, our proposal is to find new configurations, {which asymptotically yields the metric (\ref{eq:Lifshitz})},  by studying Critical Gravity (\ref{eq:Squad}) with the added contribution of a dilatonic component characterized by an Abelian field denoted as $A_{\mu}=A_{\mu}(r)dt$ and {the scalar field} $\psi=\psi(r)$, namely
}

\begin{eqnarray}
S_{\psi}&=&\int{d}^4x\sqrt{-g}
\left[ -\frac{1}{2}{\nabla_{\mu}\psi}{\nabla^{\mu}\psi}-\frac{1}{4} e^{\lambda \psi} F_{\mu \nu} F^{\mu \nu}\right]\nonumber\\
&=&\int
d^{4}x \sqrt{-g} \mathcal{L}_{\psi}, \label{eq:Lpsi}
\end{eqnarray}
with $F_{\mu \nu}:=\partial_{\mu} A_{\nu}-\partial_{\nu} A_{\mu}$.

{As a side note, it is interesting to note that this combined action $S_g+S_{\psi}$ of Critical Gravity enriched with a dilatonic component admits an additional interpretation as a Weyl-square correction of Einstein-Maxwell-Dilaton gravity. This is particularly compelling because Einstein-Maxwell-Dilaton gravity admits black hole solutions as well, as show in \cite{Gibbons:1987ps}.

The action $S_g + S_{\psi}$ has the following equations of motions:
\begin{subequations}\label{eq:EOM_g_psi}
\begin{eqnarray}
&&E_{\mu\nu} \equiv G_{\mu\nu}+\Lambda g_{\mu\nu}+K_{\mu\nu}^{C.G.}-\kappa T_{\mu\nu}^{\psi}=0, \label{eq:Einstein_psi}\\
&&\nabla_{\mu}(e^{\lambda\psi}F^{\mu\nu})=0, \label{eq:Maxwell}\\
&&\Box \psi -\frac{1}{4}\lambda e^{\lambda\psi} F^{\alpha\beta}F_{\alpha\beta}=0\label{eq:psi},
\end{eqnarray}
\end{subequations}
where $K_{\mu\nu}^{C.G.}$ and $T_{\mu\nu}^{\psi}$ are
\begin{eqnarray*}
K_{\mu\nu}^{C.G.}&=&
 2\beta_2 \Bigl(R_{\mu\rho}R_{\nu}^{\, \rho}-\frac{1}{4}R^{\rho\sigma}R_{\rho\sigma}g_{\mu\nu}\Bigr)\nonumber \\
&+&2 \beta_1 R \Bigl(R_{\mu\nu}-\frac{1}{4} R g_{\mu\nu}\Bigr)
+\beta_2  \Bigl(\Box R_{\mu\nu} + \frac{1}{2} \Box R g_{\mu\nu}\nonumber \\
&-& 2 \nabla_{\rho}\nabla_{\left(\mu\right.}R_{\left.\nu\right)}^{\,  \rho} \Bigr)
+2\beta_1 (g_{\mu\nu}\Box R-\nabla_{\mu}\nabla_{\nu} R),  \\
T_{\mu\nu}^{\psi}&=&  \nabla_{\mu}\psi \nabla_{\nu} \psi-\frac{1}{2} g_{\mu \nu}  \nabla_{\alpha} \psi \nabla^{\alpha} \psi \nonumber\\
&+& e^{\lambda\psi} \Bigl[F_{\mu\sigma}F_{\nu}^{\, \sigma} -\frac{1}{4} F^{\alpha\beta}F_{\alpha\beta}g_{\mu\nu} \Bigr],
\end{eqnarray*}
with $\beta_1,\, \beta_2$ given by (\ref{relations}), and it supports the Lifshitz spacetime (\ref{eq:Lifshitz}) only if the following conditions are met
\begin{subequations}\label{eq:asymptotic}
\begin{eqnarray}\label{eq:Lparam}
e^{\psi} &=&\mu r^{\sqrt{\frac{2\left( z-1 \right)  \left( l^2\Lambda -z^{2}+4z\right) }{\kappa l^2\Lambda}}}\label{eq:psi}
,
\label{eq:psifunction}\\\nonumber\\
F_{rt} &=& \sqrt{{\frac { \left( z-1 \right)  \left( {l}^{2}\Lambda
 -{z}^{2}+4\,z\right)  \left( z+2 \right) }{\kappa\,\Lambda\,\mu^{\lambda}}}
}\,\left(\frac{r}{l}\right)^{z+1}\label{eq:maxwelltensor}\\
\lambda&=&-\sqrt{{\frac {8 \kappa\,{l}^{2}\Lambda}{ \left( z-1 \right)  \left({l}^{2}\Lambda -{z}^{
2}+4\,z \right) }}
},\label{eq:lambda}
\\
P(\Lambda,z)&\equiv& 2\,{\Lambda}^{2} {l}^{4} + \left( 2+z \right)  \left( z+1
 \right) \Lambda {l}^{2} -z (z-1)(z-4)\nonumber \\
 &=&0, \label{eq:Lambda}
\end{eqnarray}
\end{subequations}
where $\mu$ is a positive integration constant. \textcolor{black}{It is crucial to note that, as was shown in  \cite{Taylor:2008tg,Tarrio:2011de} and \cite{BravoGaete:2017dso}, the inclusion of  the scalar field $\psi$ (\ref{eq:psi}) as well as  $A_{\mu}$ (\ref{eq:maxwelltensor}) allow Critical Gravity to sustain the Lifshitz space-time (\ref{eq:Lifshitz}) through the solution (\ref{eq:asymptotic}) for $z \neq 1$. Nevertheless, at the moment of exploring asymptotically Lifshitz black holes, these contributions do not appear in the thermodynamical relations.}

As exemplified in {\cite{Dereli:1986cm,Gibbons:2004uw,Anabalon:2009kq,Ett:2010by,Ayon-Beato:2014wla}}, a good starting point to try to find black hole configurations is through a Kerr-Schild transformation \cite{KerrSchild} of the spacetime that we want the black hole to asymptote, namely, the Lifshitz spacetime (\ref{eq:Lifshitz}).{ In this work, we will also use this technique  as a starting point to find new black hole solutions in section \ref{section:setup}. In this same section, we will establish the set up of the problem, present the equations of motion and find their solutions; in section \ref{section:thermodynamics} we will study the thermodynamics of said configurations.} Finally, in section \ref{section:conclusions} we will discuss our results and present our conclusions.

\section{Set up of the solution}\label{section:setup}
As mentioned at the end of the introduction, we will start our analysis by performing a Kerr-Schild transformation on the seed metric (\ref{eq:Lifshitz}), satisfying the conditions (\ref{eq:asymptotic}).
In our case, a Kerr-Schild transformation from the seed metric $ds_L^2$ can be expressed as:
\begin{equation}
ds^2=ds_L^2+\left(\frac{r}{l}\right)^{2z}{h}(r) k \otimes k,\label{eq:1KS}
\end{equation}
where $k$ is a null geodesic vector and $h(r)$ is a scalar function. Since we are not considering angular momentum, an appropriate choice for $k$ is
\begin{equation}
k=dt+\left(\frac{l}{r}\right)^{z+1}dr.
\end{equation}
In order to emphasise the static nature of our metric, we redefine the time coordinate as
\begin{equation}
  t \mapsto t+\int \frac{l^{z+1}}{r^{z+1}} \frac{h(r)dr}{1-h(r)}, \label{eq:2KS}
\end{equation}
which yields to:
\textcolor{black}{\begin{eqnarray}\label{eq:metric}
ds^2&=&-\frac{r^{2z}}{l^{2z}} \left[1-h(r)\right] dt^2+\frac{l^2}{r^2} \frac{dr^2}{\left[1-h(r)\right]} +\frac{r^2}{l^2}\sum_{i=1}^{2}dx_i^2.\nonumber\\
\end{eqnarray}}
In what follows, eq. (\ref{eq:metric}) will be our ansatz for the black hole metric together with the conditions (\ref{eq:asymptotic}). We also impose $\lim_{r \rightarrow
+\infty} h(r)=0$ to ensure an asymptotically Lifshitz behavior.
\\
If one considers eq. (\ref{eq:psi}) together with the conditions (\ref{eq:asymptotic}), it becomes:
\begin{equation*}
-\frac{1}{l^3}\sqrt{\frac{2(z-1)(l^2\Lambda-z^2+4z)}{\kappa l^2 \Lambda} }
 \left[ r\, h' + \left( 2+z \right)h   \right] =0.
\end{equation*}
Let us notice that equating the second factor to zero is equivalent to solving a Cauchy-Euler linear differential equation for $h(r)$. This fixes the form of $h(r)$ to
\begin{equation}\label{eq:h(r)}
  h(r)=\left(\frac{r_h}{r}\right)^{z+2},
\end{equation}
where $r_h$ is an integration constant that is related to the location of the event horizon.
Notice that another way to satisfy the equation of motion would be to consider $\sqrt{\frac{2(z-1)(l^2\Lambda-z^2+4z)}{\kappa l^2 \Lambda} }=0$. {However,} this is not permitted due to eq. (\ref{eq:asymptotic}), as this would imply the elimination of the dilatonic contribution.\\
As a result from imposing the ansatz (\ref{eq:metric}) with the conditions (\ref{eq:asymptotic}) and (\ref{eq:h(r)}) in the equations of motion (\ref{eq:EOM_g_psi}), one can find a black hole solution for $z=4$ that reads:
\begin{eqnarray}\label{eq:sol_z4}
ds^2 &=&-\frac{r^{8}}{l^{8}}\left[1-h(r)\right]dt^2 +\frac{l^2 dr^2}{r^2[1-h(r)]}
+\frac{r^2}{l^2}\sum_{i=1}^{2} dx_{i}^{2},\nonumber\\
h(r)&=&\left(\frac{r_h}{r}\right)^{6},\qquad e^\psi =\mu r^{\sqrt{\frac{6}{\kappa}}},\\
F_{rt} &=& \sqrt{\frac{18\, l^2}{\kappa \mu^{\lambda}}}\left(\frac{r}{l}\right)^{5},\qquad \Lambda=-\frac{15}{l^2}, \qquad \lambda=-\sqrt{\frac{8 \kappa}{3}}. \nonumber
\end{eqnarray}
\textcolor{black}{Just for completeness, another branch to explore Lifshitz black holes solutions is to consider the case without the constraints (\ref{relations}) together with a matter source characterized by a linear Maxwell field $F_{\mu \nu} F^{\mu \nu}$, this is, for our situation, the action (\ref{eq:Lpsi}) without a scalar field ($\psi=0$). Following the above steps,  the Lifshitz metric (\ref{eq:Lifshitz}) is supported if $\beta_1$ and $\beta_2$ are related as \cite{Ayon-Beato:2010vyw}
$$\beta_1=\frac{-2\beta_2 z^2+l^2-4 \beta_2}{4(z^2+2z+3)},$$
while that the cosmological constant takes the form
$$\Lambda=-\frac{z^2+2z+3}{2 l^2},$$
without matter source. Then, through the Kerr-Schild transformation (\ref{eq:1KS}) with the metric (\ref{eq:metric}) the Einstein equations yield to two solutions, one of them corresponds to the four-dimensional Lifshitz black hole configuration in vacuum when general quadratic corrections are considered and $z=6$, found in \cite{Ayon-Beato:2010vyw}, which reads
\begin{eqnarray*}
ds^2 &=&-\frac{r^{12}}{l^{12}}\left[1-\left(\frac{r_h}{r}\right)^{4}\right]dt^2 +\frac{l^2}{r^2} \left[1-\left(\frac{r_h}{r}\right)^{4}\right]^{-1} dr^2\\
&+&\frac{r^2}{l^2}\sum_{i=1}^{2} dx_{i}^{2},\\
\beta_1&=&-\frac{9l^2}{64},\qquad \beta_2=\frac{25 l^2}{64},\qquad \Lambda=-\frac{51}{2l^2}.
\end{eqnarray*}
On the other hand, and as was shown in \cite{Bravo-Gaete:2015xea}, it is possible to obtain a four-dimensional  charged Lifshitz black hole with dynamical exponent $z = 3$, given by
\begin{eqnarray*}
ds^2 &=&-\frac{r^{6}}{l^{6}}\left[1-\left(\frac{r_h}{r}\right)^{4}\right]dt^2 +\frac{l^2}{r^2} \left[1-\left(\frac{r_h}{r}\right)^{4}\right]^{-1} dr^2\\
&+&\frac{r^2}{l^2}\sum_{i=1}^{2} dx_{i}^{2},\\
F_{rt}&=&2 \sqrt{-\frac{3 \beta_2}{\kappa}} \frac{r_h^2}{l^{4}},\\
\beta_1 &=& -\frac{11 \beta_2}{36}+\frac{l^2}{72},\qquad \Lambda = -\frac{9}{l^2},
\end{eqnarray*}
where in order to obtain a real solution, the constant $\beta_2$ must be negative.
}

\textcolor{black}{While both of these solutions are of general interest, let us continue to focus on the case of Critical Gravity (which entails constraining the values of $\beta_1$ and $\beta_2$ as in eq. (\ref{relations})). So far, we have described solution (\ref{eq:sol_z4}) for the theory of Critical Gravity with a dilatonic contribution. The solution (\ref{eq:sol_z4})}, however correct, does not fully answer the questions that we asked during the introduction, since the solution of the equations of motion restricts the value of the dynamical exponent to be $z=4$. As such, while it represents an asymptotically Lifshitz black hole supported by Critical Gravity (albeit through the addition of a dilatonic contribution), it cannot be used to recover the Schwarzchild-AdS limit, since it has a fixed value of $z$.
Nevertheless, it has been shown in \cite{Correa:2014ika,Ayon-Beato:2015jga,Ayon-Beato:2019kmz,Bravo-Gaete:2020ftn}, that the addition of a non-minimally coupled scalar field can be helpful to enrich the equations of motion. In the interest of liberating $z$ from its previous constraints, we will too consider the addition of an action $S_{\Phi}$ including a non-minimally coupled scalar field, that is:
\begin{eqnarray}\label{eq:action}
  S = S_g+S_{\psi}+S_{\Phi},
\end{eqnarray}
where
\begin{eqnarray}
S_{\Phi}&=&\int{d}^4x\sqrt{-g}\Biggl[
-\frac{1}{2}\nabla_{\mu}\Phi\nabla^{\mu}\Phi-\frac{\xi}{2}R\Phi^2-U(\Phi)\Biggr]\nonumber\\
&=&\int
d^{4}x \sqrt{-g} \mathcal{L}_{\Phi},\label{eq:LPhi}
\end{eqnarray}
and its equations of motion are:
\begin{subequations}\label{eq:EOM}
\begin{eqnarray}
&&E_{\mu\nu}-\kappa T_{\mu\nu}^{\Phi}=0, \label{eq:Einstein_psi_phi}\\
&& \Box \Phi-\xi R\Phi=\frac{dU(\Phi)}{d\Phi}\label{eq:phi},
\end{eqnarray}
\end{subequations}
together with equations (\ref{eq:Maxwell}) and (\ref{eq:psi}), where $T_{\mu\nu}^{\Phi}$ is
\begin{eqnarray}
T_{\mu\nu}^{\Phi}&=& \nabla_{\mu}\Phi\nabla_{\nu}\Phi - g_{\mu\nu}\Bigl[\frac{1}{2}\nabla_{\sigma}\Phi\nabla^{\sigma}\Phi +U(\Phi)\Bigr]\nonumber \\
&+& \xi(g_{\mu\nu}\Box -\nabla_{\mu\nu}+G_{\mu\nu} )\Phi^2,
\end{eqnarray}
and $E_{\mu\nu}$ can be found in (\ref{eq:Einstein_psi}).
Let us remark that since the fields $\psi$ and $\Phi$ do not interact in the action, the condition imposed in the form of $h(r)$ in (\ref{eq:h(r)}) still holds.
{Indeed, using this starting point and the asymptotic conditions to recover the Lifshitz spacetime when {$r\rightarrow +\infty$}, we find the Lifshitz black hole:
\begin{subequations}\label{solution}
\begin{eqnarray}
ds^2 &=&-\frac{r^{2z}}{\l^{2z}}[1-h(r)]dt^2 +\frac{\l^2 dr^2}{r^2 [1-h(r)]}
+\frac{r^2}{\l^2}\sum_{i=1}^{2} dx_{i}^{2},\nonumber \\
h(r)&=&\left(\frac{r_h}{r}\right)^{z+2},\label{eq:fbh}\\
\Phi(r)&=&\sqrt{-{\frac {24(z-1)}{\kappa\,{l}^{2}\Lambda}}}\left(\frac{r_h}{r}\right)^{\frac{z+2}{2}},\label{eq:phibh}
\\
e^{\psi} &=&\mu r^{\sqrt{\frac{2 \left( z-1 \right)  \left( l^{2}\Lambda -{z}^{2}+4\,z\right) }{\kappa l^{2}\Lambda}}}
,
\\\nonumber\\
F_{rt} &=& \sqrt{{\frac { \left( z-1 \right)  \left( {l}^{2}\Lambda
 -{z}^{2}+4\,z\right)  \left( z+2 \right) }{\kappa\,\Lambda\,\mu^{\lambda}}}
}\left(\frac{r}{l}\right)^{z+1},\label{eq:Frtphi}\\
\lambda&=&-\sqrt{{\frac {8 \kappa\,{l}^{2}\Lambda}{ \left( z-1 \right)  \left({l}^{2}\Lambda -{z}^{
2}+4\,z \right) }}
},
\end{eqnarray}
where $\Lambda$ is constraint to obey the restriction
\begin{eqnarray}
{P(z,\Lambda l^{2})}&\equiv& {2\,({\Lambda} {l}^{2})^2} + \left( 2+z \right)  \left( z+1
 \right) \Lambda {l}^{2} \nonumber \\
&-&z (z-1)(z-4)=0,
\end{eqnarray}
the non-minimal coupling parameter is given by
\begin{equation}
\xi=\frac{1}{6},\label{eq:xi}
\end{equation}
and the potential $U(\Phi)$ is tied as follows
\begin{eqnarray}\label{eq:potential}
U(\Phi) &=& \frac{z(z-4)}{24 l^2} \Phi^2+ \frac{\kappa \Lambda(3z^2+8z+16)}{1152\, (z-1)}\Phi^4.
\end{eqnarray}
\end{subequations}
Let us note that the scalar field $\psi$, the Maxwell field $F_{rt}$, the constant $\lambda$ together with the relation of the cosmological constant $\Lambda$ and the dynamical exponent $z$ are the ones expressed in (\ref{eq:Lparam})-(\ref{eq:Lambda}), and that arise from demanding that the Lifshitz spacetime is recovered asymptotically.}

In order to ensure the reality of $\psi$, $\Phi$, $F_{rt}$ and $\lambda$, for $\kappa>0$, this solution must be defined in the ranges {shown} in the Table \ref{tabla1}.
\begin{table}[h!]
\begin{tabular}{|c|c|}
\hline
Range of $z$  &  Range of $\Lambda l^2$ \\
\hline \hline
 $-2<z<0$ &  $0<\Lambda l^2<{z(z-4)}$\\  [1ex]
\hline \hline
 $1<z<4$  & $\Lambda l^2<{z(z-4)}$ \\  [1ex]
\hline \hline
$z\geq 4$  & $\Lambda l^2<0$ \\  [1ex]
\hline
\end{tabular}
\caption{\label{tabla1}Range of possibilities for the dynamical
exponent $z$ and $\Lambda l^2$ permitting the reality of $\psi$, $\Phi$, $F_{rt}$ and $\lambda$.}
\end{table}%

However, there is one more condition imposed on the solution given by eq.(\ref{eq:Lambda}), which we express through Figure \ref{fig:P_Lambda} for clarity. Let us notice that Table \ref{tabla1} imposes a particular sign for {$\Lambda l^{2}$} for different intervals of $z$. The only interval in which these conditions are compatible with Figure \ref{fig:P_Lambda} is $z\in(1,4)$, {represented via the intersection between the curve $P(z,\Lambda l^{2})$=0 and the region ${\cal{R}}$}. As a result, the black hole configuration given by (\ref{solution}) is only valid, \emph{$\grave{a}$ priori}, on this range. Moreover, we note that in this interval $z\in(1,4)$, the potential $U(\Phi)$ is positive, that is, the potential is bounded from below.

\begin{figure}[h!]
  \centering
    \includegraphics[scale=.18]{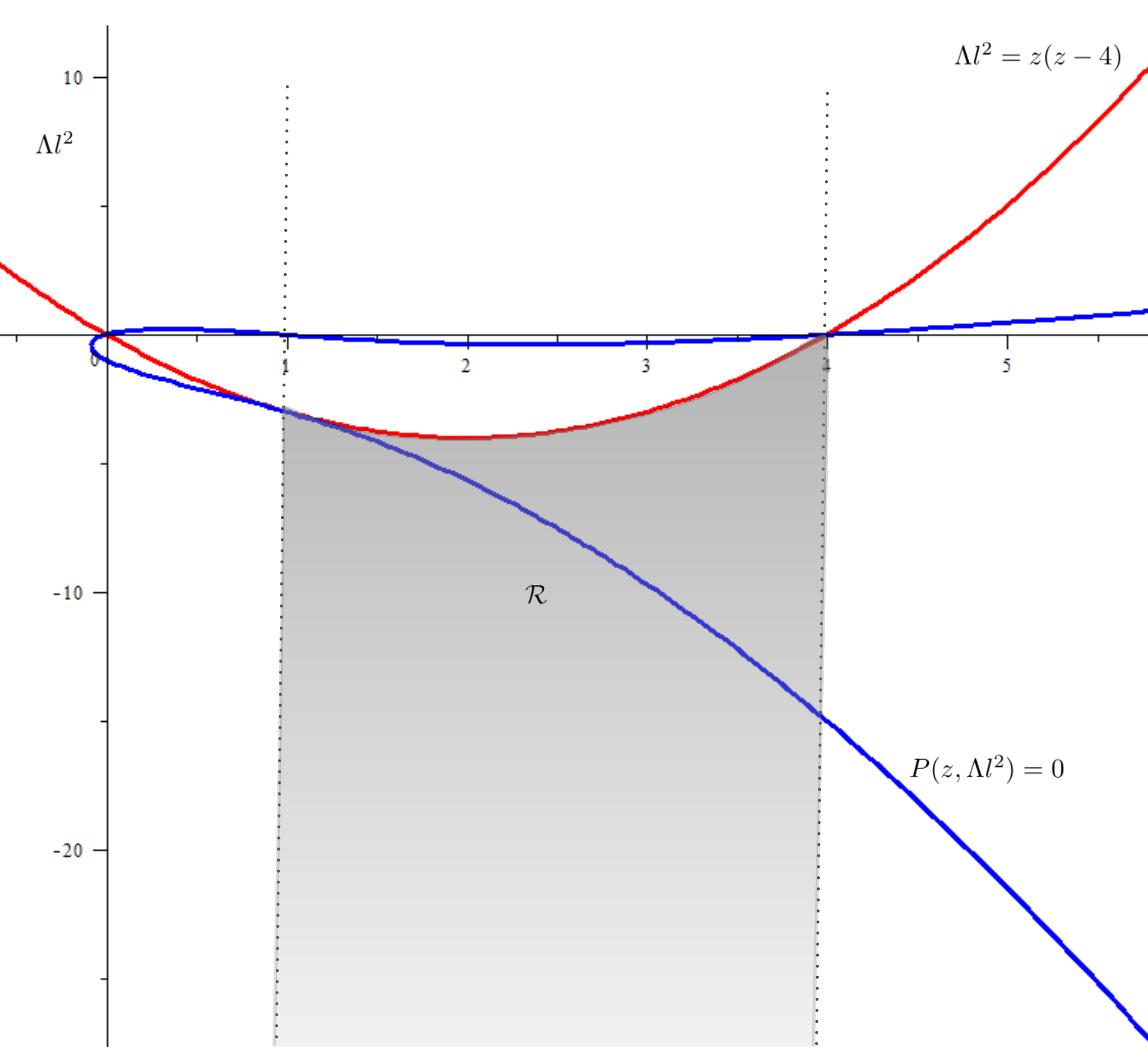}
      \caption{Representation of {$P(z,\Lambda l^2)=0$} and the polynomial $\Lambda l^2=z(z-4)$. {Together with the above, ${\cal{R}}=\{(z,\Lambda l^{2}) \in \mathbb{R}^{2} / 1<z<4 \mbox{ and } \Lambda l^{2}<z(z-4) \}$}}
  \label{fig:P_Lambda}
\end{figure}

A few more comments can be made {regarding} this solution. First of all, in the limit case $z \rightarrow 1$, although the coupling constant $\lambda$ blows up, the dilaton $e^{\lambda \psi} F_{\mu \nu} F^{\mu \nu}$ goes to zero, the scalar fields $\psi$, $\Phi$ as well as the potential { $U(\Phi)$ } vanish identically and one ends  up with the Schwarzschild-AdS solution with planar-base manifold and the cosmological constant $\Lambda=-3/l^2$. Curiously enough, for this hairy anisotropic geometry, the nonminimal coupling parameter must be the four-dimensional conformal $\xi=1/6$.

Another interesting particular case would be when $z=4$. In this scenario, even though $\Phi(r)$, $\xi$ and $U(\Phi)$ do not vanish, we find that $T_{\mu\nu}^{\Phi}=0$, leading to the solution  (\ref{eq:sol_z4}) together with
\begin{eqnarray}\label{eq:sol_z_4_phi}
\Phi(r)&=&\sqrt{\frac{24 }{5 \kappa}}\left(\frac{r_h}{r}\right)^{3},\qquad U(\Phi)=-\frac{5 \kappa}{12 l^2} \Phi^4,
\end{eqnarray}
and the non-minimal coupling parameter (\ref{eq:xi}). Let us note that the configuration given previously in (\ref{eq:sol_z4}) and the configuration above can be seen as two solutions from the same theory $S_g+S_{\psi}+S_{\Phi}$. The first one would correspond to a solution (\ref{eq:sol_z4}) with $\Phi=0$, while the second one would correspond to a solution (\ref{eq:sol_z4}) with $\Phi$ given by (\ref{eq:sol_z_4_phi}). { An} interesting interpretation of this solution (\ref{eq:sol_z_4_phi}) is that it can be seen as a stealth-like configuration, more specifically, the field $\Phi$ can be seen as a stealth field overflying the black hole (\ref{eq:sol_z4}).

\textcolor{black}{Here it is important to note that the above \emph{stealth}-like configuration is not a \emph{stealth} in a traditional way. A \emph{stealth} field is a non-trivial field whose associated energy-momentum tensor is zero, that is, it has no gravitational backreaction. Pioneering work on this front can be found in \cite{AyonBeato:2004ig,AyonBeato:2005tu,Ayon-Beato:2015qfa}. Following the traditional definition, in our case, would imply that there exist fields, such that the whole energy-momentum tensor vanishes non-trivially (independently from the gravitational side), that is:
$$ G_{\mu\nu}+\Lambda g_{\mu\nu}+K_{\mu\nu}^{C.G.}=0=\kappa \left(T_{\mu\nu}^{\psi}+ T_{\mu\nu}^{\Phi}\right).$$
This, however, is not possible, since it would imply that the left-hand side can vanish identically on its own, while we know that Critical Gravity cannot support the Lifshitz spacetime (\ref{eq:Lifshitz})  on its own.}

\textcolor{black}{In order to reinforce the above, we start with the metric (\ref{eq:Lifshitz}), and the scalar field $\Phi$ as well as the non-minimal coupling parameter $\xi$ are given by \cite{Ayon-Beato:2015qfa}
\begin{equation}\label{eq:s-g}
\Phi(r)=\frac{\sqrt{\Phi_0}}{r^{\frac{z+2}{2}}},\qquad \xi=\frac{(2+z)^2}{8(z^2+2z+3)},
\end{equation}
where $\Phi_0$ is an integration constant, and $U(\Phi)=0$. Next, from imposing $E_{\mu\nu}=0$ (eq. \ref{eq:Einstein_psi}), we find that the solution (\ref{eq:Lparam})-(\ref{eq:Lambda}), is the most general configuration on the line element (\ref{eq:Lifshitz}).
Nevertheless, if one wants to find asymptotically Lifshitz black holes through a Kerr-Schild transformation (\ref{eq:1KS})-(\ref{eq:2KS}), the combination of equations of motion $(t,t)$-$(r,r)$ from  (\ref{eq:Einstein_psi_phi})  imposes that $z=4$. This, in turn, fixes the coupling $\xi$ to be the conformal one ($\xi=1/6$) and the cosmological constant to be $\Lambda=-15/l^2$ from (\ref{eq:xi}) and (\ref{eq:sol_z4}) respectively. Finally the $(t,t)$ (or $(x_i,x_i)$) component of  (\ref{eq:Einstein_psi_phi}) yields an equation proportional to
\begin{equation}\label{eq:E-q}
\frac{2\kappa \Phi_0 r_h^6}{r^{12} l^2}=0,
\end{equation}
 on one side we recover a particular solution from (\ref{eq:s-g}) if $r_h=0$, and on the other side the scalar field $\Phi$ vanishes trivially, and the configuration (\ref{eq:Lparam})-(\ref{eq:Lambda}) is obtained for $z=4$. }

\textcolor{black}{It is worth pointing out that the above analysis does not consider the presence of the potential $U(\Phi)$, which vanishes when we consider the solution of $T_{\mu \nu}^{\Phi}=0$ (\ref{eq:s-g}) on the Lifshitz metric  (\ref{eq:Lifshitz}). Nevertheless, adding a potential $U(\Phi) \propto \Phi^4$ for the above study, the last Einstein equation (\ref{eq:E-q}) is compensated via the presence of $U(\Phi)$, and the solution found previously in (\ref{eq:sol_z_4_phi}) is recovered, where the integrations constants $\Phi_0$ and $r_h$ are related.}

In order to further distinguish these two solutions, a plan of action is to analyze their thermodynamics. Taking this as a motivation, in the following section we study their thermodynamical quantities.

\section{Thermodynamics of the solutions}\label{section:thermodynamics}

Regarding the thermodynamical properties of the configuration (\ref{solution}),
the entropy of the Lifshitz black hole solutions can be calculated
by using Wald's formula \cite{Wald:1993nt,Iyer:1994ys}, which can be read as
\begin{eqnarray}
 \mathcal{S}_{W} &=& \left.- 2 \pi \, \Omega_{2} \, \left(\frac{r_h}{\l}\right)^{2} \,
 \Big( P^{\mu \nu \sigma \rho} \, \varepsilon_{\mu \nu} \,  \varepsilon_{\sigma \rho} \Big)\right|_{r = r_{h}}\nonumber\\
 &=&{\frac {2\,\pi \,\Omega_{{2}} \left( {l}^{2}\Lambda-{z}^{2}+4\,z \right)}{\Lambda\,\kappa\,{l}^{2}}}
\left(\frac{r_h}{l}\right)^{2}. \label{eq:Sw}
\end{eqnarray}
{Here $\Omega_{2}$ represents the finite contribution of the
$2$-dimensional integration over the planar variables, $\varepsilon_{\mu \nu}$ is the binormal vector that follows from the timelike Killing vector {$\partial_{t}=k^{\mu}\partial_{\mu}$}, which becomes null at the event horizon $r_h$ and reads
$$\varepsilon_{\mu \nu}=-\varepsilon_{\nu \mu}:=\frac{\nabla_{\mu} k_{\nu}}{\sqrt{-\frac{1}{2}\left(\nabla_{\mu} k_{\nu}\right)\left(\nabla^{\mu} k^{\nu}\right)}},$$
while
\begin{eqnarray*}
P^{\alpha \beta \gamma \delta}&=&\frac{\delta \mathcal{L}}{\delta
R_{\alpha \beta \gamma \delta}}\nonumber\\
&=&\frac{1}{4}\,\left(\frac{1}{\kappa}-\xi
\Phi^{2}\right)\Big(g^{\alpha\gamma}g^{\beta\delta}-g^{\alpha\delta}g^{\beta\gamma}\Big)
\nonumber\\
&+&\frac{\beta_1}{2 \kappa}\, R \left(g^{\alpha \gamma } g^{\beta
\delta } -g^{\alpha \delta } g^{\beta \gamma
}\right)\nonumber\\
&+&\frac{\beta_{2}}{4 \kappa}\, \left(g^{\beta \delta } R^{\alpha
\gamma }-g^{\beta \gamma } R^{\alpha \delta } -g^{\alpha \delta }
R^{\beta \gamma }+g^{\alpha \gamma } R^{\beta \delta
}\right),
\end{eqnarray*}
where are $\beta_1$ and $\beta_2$ tied as (\ref{relations}), and the Hawking temperature $T$ is given by
\begin{eqnarray}
 T&=&\left.\frac{\sqrt{-\frac{1}{2}\left(\nabla_{\mu} k_{\nu}\right)\left(\nabla^{\mu} k^{\nu}\right)}}{2\pi} \right|_{r=r_h}\nonumber\\
&=&\frac{1}{4\,\pi} \frac{r_{h}^{z+1}}{\l^{z+1}}
 f^{\prime}(r_{h})= \frac{(z+2)}{4 \pi l}\left(\frac{r_h}{l}\right)^{z}. \label{eq:T}
\end{eqnarray}}

In order to obtain the mass of these Lifshitz black hole
solutions,
we follow the generalized method proposed in \cite{Kim:2013zha,Gim:2014nba},
based on the off-shell extension of the Abbott-Deser-Tekin (ADT) formalism \cite{Abbott:1981ff,Deser:2002rt,Deser:2002jk}. {This technique has shown to be useful to compute the mass of anisotropic black hole solutions  \cite{Herrera-Aguilar:2020iti}, their solitonic counterparts \cite{Bravo-Gaete:2020ftn,Ayon-Beato:2019kmz,Bravo-Gaete:2015iwa}, and even the angular momentum for spinning configurations \cite{BravoGaete:2017dso}.} Following
the notation of these previous works, the quasilocal conserved charge can be read as
\begin{equation}
\label{eq:5} \mathcal{M}(k)=\int_{\cal B}
d^{2}x_{\mu\nu}\Big(\Delta K^{\mu\nu}(k)-2k^{[\mu}\int^1_0ds~
\Theta^{\nu]}(k\, |\, s\mathcal{M})\Big),
\end{equation}
where $\Delta K^{\mu\nu}(k) \equiv
K^{\mu\nu}_{s=1}(k)-K^{\mu\nu}_{s=0}(k)$, $K^{\mu\nu}_{s=0}$
means the Noether potential of the vacuum solution,
$\Theta^{\nu}$ denotes a surface term and $k^{t}=(1,0,0,0)$ is the timelike Killing vector field. In our case, the surface term
and the tensor $K^{\mu\nu}$ are given by
\begin{align}
\Theta^\mu &=2\sqrt{-g}\Big[P^{\mu(\alpha
\beta)\gamma}\nabla_\gamma\delta g_{\alpha\beta}
-\delta g_{\alpha\beta}\nabla_\gamma P^{\mu(\alpha\beta)\gamma}\nonumber\\
&+\frac{1}{2}\,\frac{\delta \mathcal{L}}{\delta \left(\partial_{\mu}\Phi\right)}\delta \Phi+\frac{1}{2}\,\frac{\delta \mathcal{L}}{\delta \left(\partial_{\mu}\psi\right)}\delta \psi
\nonumber\\
&+\frac{1}{2}\,\frac{\delta \mathcal{L}}{\delta \left(\partial_{\mu} A_{\nu}\right)}\delta A_{\nu}\Big], \label{eq:theta}\\
K^{\mu\nu} &=\sqrt{-g}\,\Big[2P^{\mu\nu\rho\sigma}\nabla_\rho k_\sigma
-4k_\sigma\nabla_\rho P^{\mu\nu\rho\sigma}\nonumber\\
&-\frac{\delta \mathcal{L}}{\delta \left(\partial_{\mu} A_{\nu}\right)} k^{\sigma} A_{\sigma}\Big] \label{eq:K},
\end{align}
\textcolor{black}{where, as before, and following the steps performed in \cite{Taylor:2008tg,Tarrio:2011de} and \cite{BravoGaete:2017dso}, is important to note that the introduction of the scalar field $\psi$ (\ref{eq:psi}) and the vector potential $A_{\mu}$ (\ref{eq:maxwelltensor}) allow us to support the Lifshitz space-time (\ref{eq:Lifshitz}) for $z \neq 1$, and the only integration constant associated to the thermodynamical parameters corresponds to $r_h$ (the location of the event horizon), implying that the only thermodynamical parameters arising are the mass $\mathcal{M}_{\mathrm{bh}}$ and the entropy $\mathcal{S}_{W}$.
}

With all these ingredients, we calculate the mass $\mathcal{M}_{\mathrm{bh}}$ of the black hole, and obtain
\begin{equation}\label{eq:mass_phi}
\mathcal{M}_{\mathrm{bh}}={\frac {\Omega_{{2}} \left( {l}^{2}\Lambda-{z}^{2}+4\,z \right)}{\Lambda\,\kappa\,{l}^{3}}}
\left(\frac{r_h}{l}\right)^{z+2},
\end{equation}
which allows us to verify that the First Law of black hole thermodynamics
\begin{eqnarray}
d\mathcal{M}_{\mathrm{bh}}=T d\mathcal{S}_{W},\label{firstlaw}
\end{eqnarray}
holds. An interesting detail to explore is wether our new configurations fulfill a Smarr relation \cite{Smarr:1972kt}. {Let us note that, rewriting the mass (\ref{eq:mass_phi})  as a function of the extensive quantitative $\mathcal{S}_{W}$ from (\ref{eq:Sw}) as follows
$$\mathcal{M}_{\mathrm{bh}}(\mathcal{S}_{W})=\frac{l^{z-1} \big(\mathcal{S}_{W}\big)^{\frac{z+1}{2}}}{2 \pi} \left[\frac{\Lambda \kappa}{2\,\pi \,\Omega_{{2}} \left( {l}^{2}\Lambda-{z}^{2}+4\,z \right)}\right]^{\frac{z}{2}},$$
we have
$$\mathcal{M}_{\mathrm{bh}}(\mathcal{S}_{W})=\left(\frac{2}{z+2}\right) \left(\frac{\partial \mathcal{M}_{\mathrm{bh}}}{ \partial \mathcal{S}_{W}} \right)\mathcal{S}_{W},$$
which yields to a four dimensional Smarr relation:
\begin{eqnarray}
\mathcal{M}_{\mathrm{bh}}=\left(\frac{2}{z+2}\right)T \mathcal{S}_{W}.\label{smarr}
\end{eqnarray}
For the sake of completeness, we note that under a re-scaling with a parameter $\zeta$, we have
$$\mathcal{M}_{\mathrm{bh}}(\zeta\mathcal{S}_{W})=\zeta^{\frac{z+1}{2}}\mathcal{M}_{\mathrm{bh}}(\mathcal{S}_{W}).$$}
Let us also notice that the interval of existence of the solution $z\in(1,4)$, that we established based on reality conditions is also the only one for which the Wald entropy (\ref{eq:Sw}) and the mass (\ref{eq:mass_phi})  are positive.
Two other interesting notes are in order. First, when we consider the Schwarzchild-AdS case ($z=1$ and $\Lambda=-3/l^2$), the entropy and the mass vanish trivially as expected. Secondly, when one analyzes the $z=4$ case, we have two branches holding the same Hawking temperature $$ T=\frac{3}{2 \pi l} \left(\frac{r_h}{l}\right)^{4}, $$ the first one, characterized by the solution (\ref{eq:sol_z4}) where we have
\begin{eqnarray}
\mathcal{S}^{(1)}_{W}&=&\frac {18 \pi  \Omega_2}{5 \kappa} \left(\frac{r_h}{l}\right)^2,\qquad
\mathcal{M}^{(1)}_{\mathrm{bh}}=\frac {9 \,\Omega_2}{5 \kappa l} \left(\frac{r_h}{l}\right)^6,\label{eq:entmassz4}
\end{eqnarray}
and the other one, given by solution  (\ref{eq:sol_z4}), (\ref{eq:xi}) and (\ref{eq:sol_z_4_phi}) where we obtain
\begin{eqnarray}
\mathcal{S}^{(2)}_{W}&=&\frac {2 \pi  \Omega_2}{\kappa} \left(\frac{r_h}{l}\right)^2,\qquad
\mathcal{M}^{(2)}_{\mathrm{bh}}=\frac {\Omega_2}{\kappa l} \left(\frac{r_h}{l}\right)^6.\label{eq:entmassz4phi}
\end{eqnarray}
As before, the First Law (\ref{firstlaw}) and the Smarr relation (\ref{smarr}) (with $z=4$) hold for these configurations.

Just for completeness, with the above information we are in a position to analyze this thermodynamical system where small perturbations around the equilibrium are allowed. For this, we will consider the canonical ensemble where the quantity $T$ is fixed. Concretely, we can solve $r_h$ in (\ref{eq:T}), which reads
$$r_h=l \left(\frac{4 \pi T l}{z+2}\right)^{\frac{1}{z}},$$
and the Wald entropy (\ref{eq:Sw}) as well as the mass (\ref{eq:mass_phi}) can be expressed as
\begin{eqnarray*}
\mathcal{S}_{W}={\frac {2 \pi \Omega_{2} \left( {l}^{2}\Lambda-{z}^{2}+4\,z \right) T ^{\frac{2}{z}}}{\Lambda\,\kappa\,{l}^{2}}} \left(\frac{4 \pi  l}{z+2}\right)^{\frac{2}{z}},\\
\mathcal{M}_{\mathrm{bh}}={\frac {\Omega_{2} \left( {l}^{2}\Lambda-{z}^{2}+4\,z \right) T ^{\frac{z+2}{z}}}{\Lambda\,\kappa\,{l}^{3}}} \left(\frac{4 \pi  l}{z+2}\right)^{\frac{z+2}{z}},
\end{eqnarray*}
where we can study the local thermodynamical stability through to the heat capacity $C$, given by
\begin{eqnarray*}
C=\frac{\partial \mathcal{M}_{\mathrm{bh}}}{\partial T}&=& T \left(\frac{\partial \mathcal{S}_{W}}{\partial T}\right)\\
&=&{\frac {4 \pi \Omega_{2} \left( {l}^{2}\Lambda-{z}^{2}+4\,z \right) T ^{\frac{2}{z}}}{\Lambda\,\kappa\,{l}^{2}}} \left(\frac{4 \pi  l}{z+2}\right)^{\frac{2}{z}},
\end{eqnarray*}
which is positive for $z \in (1,4]$ and $\kappa>0$, allowing us to conclude that this asymptotically Lifshitz black hole is locally thermodynamically
stable. A similar conclusion can be reached regarding the solution found previously in (\ref{eq:sol_z4}), where
$${C=\frac{\partial \mathcal{M}^{(1)}_{\mathrm{bh}}}{\partial T}=T \left(\frac{\partial \mathcal{S}^{(1)}_{W}}{\partial T}\right)=
\frac{3 \pi \Omega_{2} \sqrt{6 \pi l  T}}{5\kappa},}
$$
is always positive.
\begin{figure}[h!]
  \centering
    \includegraphics[scale=.138]{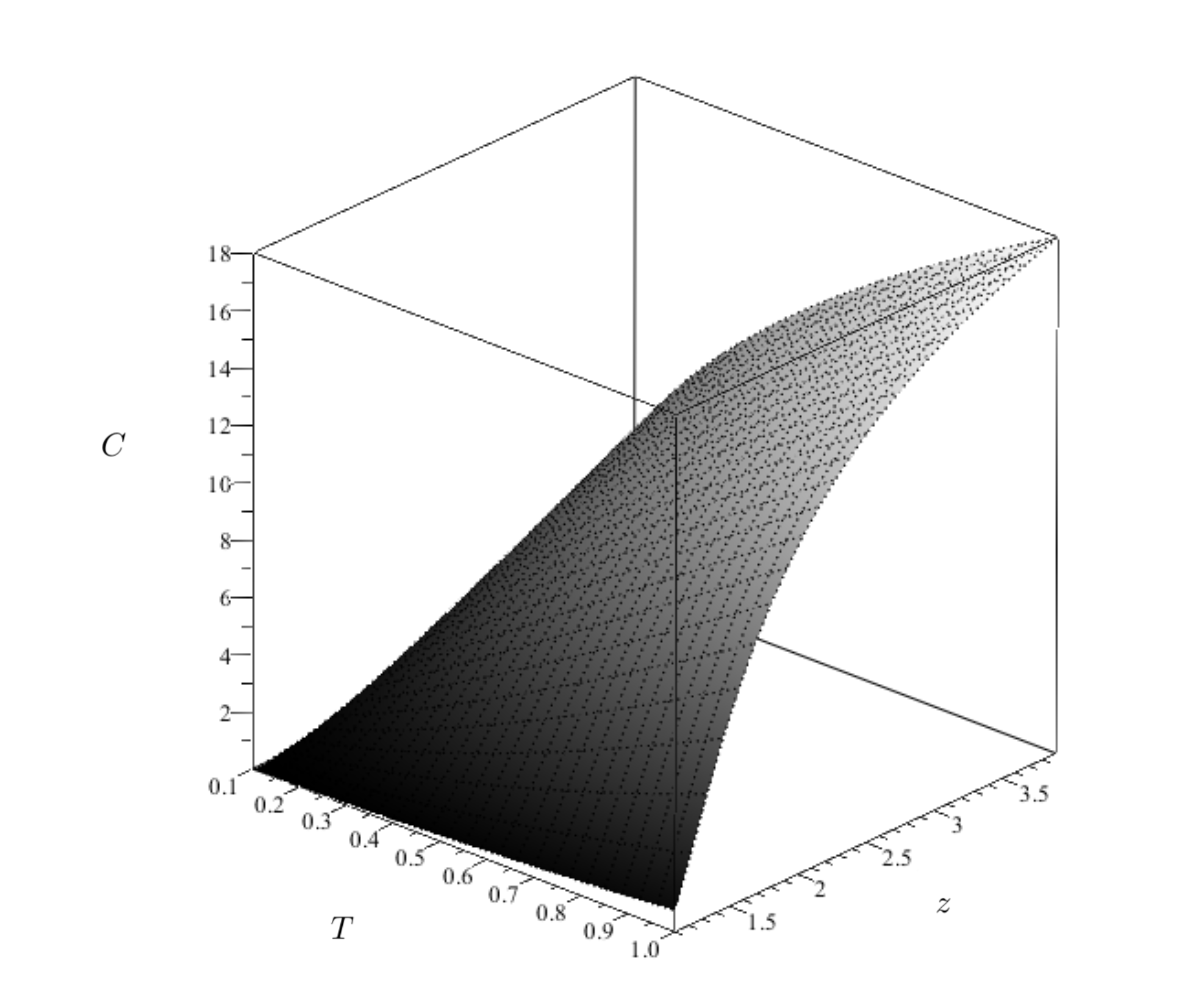}
    \includegraphics[scale=.138]{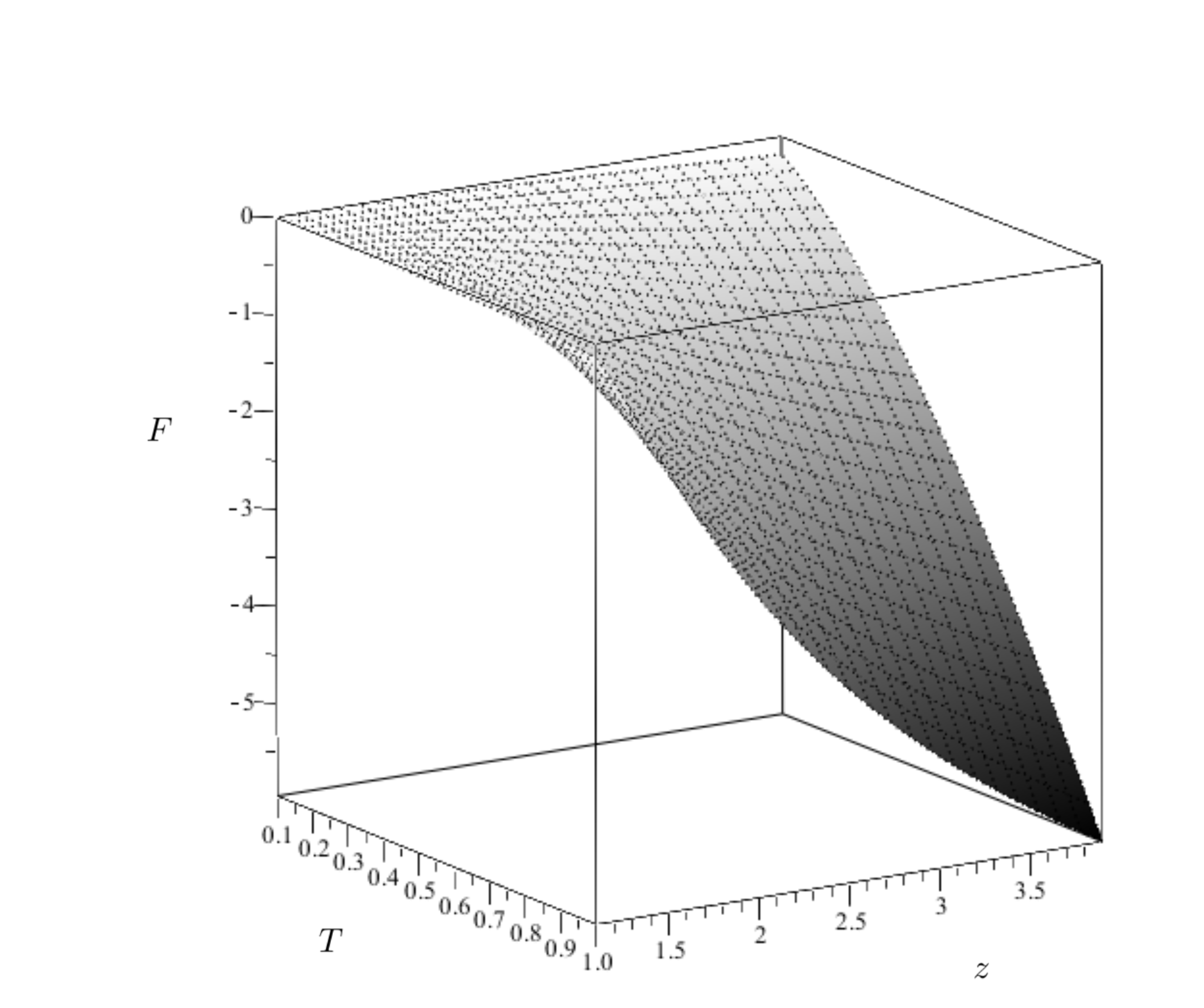}
      \caption{Upper panel: Plot of the specific heat $C$ as a function of the temperature $T$ and $z$.  Lower panel: Plot of the free energy $F$ as a function of the temperature $T$ and $z$.  For both cases, we consider $T>0$ and $z \in (1,4]$ and impose $\kappa=l=\Omega_2=1$ for simplicity.  }
  \label{fig:3d}
\end{figure}

{The global stability can be studied by using the concavity of the Helmholtz
free energy $F$ which, for the solution (\ref{solution}), reads
\begin{eqnarray}
F&=&\mathcal{M}_{\mathrm{bh}}-T \mathcal{S}_{\mathrm{bh}},\nonumber\\
&=&-\frac{\Omega_2 z (l^2 \Lambda-z^2+4z)}{2\Lambda \kappa l^3} \left(\frac{r_h}{r}\right)^{z+2},\nonumber\\
&=&-{\frac {z \Omega_{2} \left( {l}^{2}\Lambda-{z}^{2}+4\,z \right) T ^{\frac{z+2}{z}}}{2 \Lambda\,\kappa\,{l}^{3}}} \left(\frac{4 \pi  l}{z+2}\right)^{\frac{z+2}{z}},\label{gibbs-z}
\end{eqnarray}
as a function of the extensive quantity $T$. Analyzing this expression, we find that $F<0$ and $\frac{\partial^{2} F}{\partial T^{2}}<0$, which ensures the global thermodynamic stability. This analysis is analogous for the configuration (\ref{eq:sol_z4}). The general cases for the specific heat $C$ and the free energy $F$ are represented in Figure \ref{fig:3d}.
}

{Finally, given that we have a special situation of two anisotropic solutions with $z=4$, that have the same asymptotic behavior and Hawking Temperature, we can analyze the free energy $F$ between them, {where  $F^{(1)}$ (respectively  $F^{(2)}$) corresponds to the solution (\ref{eq:sol_z4}) with $\Phi=0$ (respectively the stealth solution  characterized by the combination of (\ref{eq:sol_z4}), (\ref{eq:xi}) and (\ref{eq:sol_z_4_phi})), given by}
\begin{eqnarray}
F^{(1)}&:=&\mathcal{M}^{(1)}_{\mathrm{bh}}-T \mathcal{S}^{(1)}_{\mathrm{bh}}=-\frac{4 \Omega_{2} \sqrt{6\pi l} \left(\pi T\right)^{3/2}}{5 \kappa},\\
F^{(2)}&:=&\mathcal{M}^{(2)}_{\mathrm{bh}}-T \mathcal{S}^{(2)}_{\mathrm{bh}}=
-\frac{4 \Omega_{2} \sqrt{6 \pi l} \left(\pi T\right)^{3/2}}{9 \kappa},
\end{eqnarray}
and
$$\Delta F:=F^{(1)}-F^{(2)}=-\frac{16 \Omega_{2} \sqrt{6 \pi l} \left(\pi T\right)^{3/2}}{45 \kappa}<0,$$
for $T>0$, concluding that \textcolor{black}{the Lifshitz black hole (\ref{eq:sol_z4}), is thermodynamically preferred over the configuration that supports the stealth (\ref{eq:sol_z4}), (\ref{eq:xi})- (\ref{eq:sol_z_4_phi}).} This situation is shown in Figure \ref{fig:Gibbs}.
}
\begin{figure}[h!]
  \centering
    \includegraphics[scale=.33]{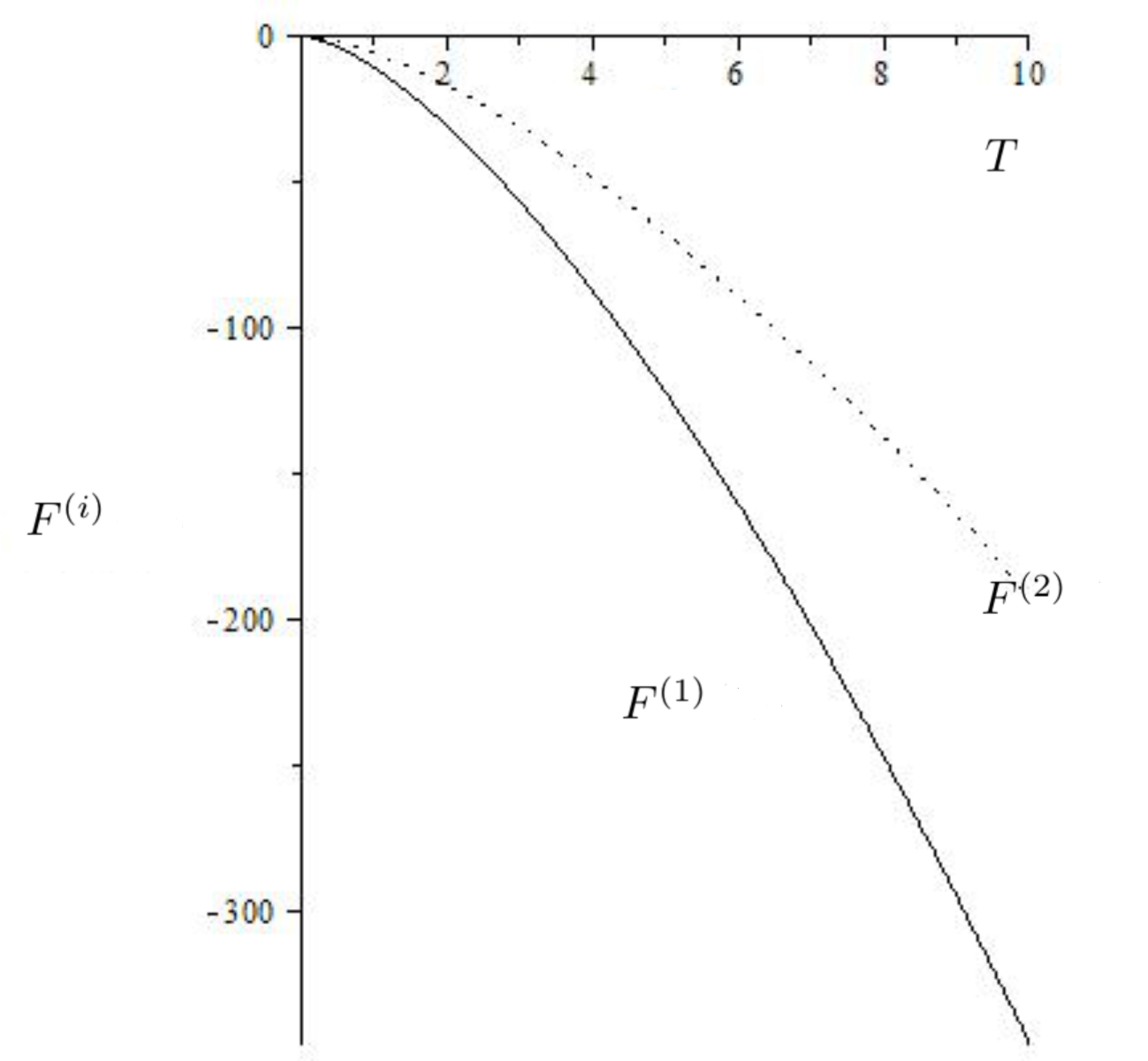}
      \caption{Free energy $F^{(1)}$ (respectively  $F^{(2)}$) in function of temperature $T$ of the anisotropic black hole configuration (\ref{eq:sol_z4}) (respectively  (\ref{eq:sol_z4}), (\ref{eq:xi}) and (\ref{eq:sol_z_4_phi})), being represented by a continuous curve (respectively by a dashed curve). Here we impose $\kappa=l=\Omega_2=1$ {for simplicity}. }
  \label{fig:Gibbs}
\end{figure}

\section{Discussion and Conclusions}\label{section:conclusions}
The aim of this work was to answer \textcolor{black}{three} questions regarding the possibility to find asymptotically Lifshitz black holes in Critical Gravity. In our effort to answer these questions, inspired by, Taylor \cite{Taylor:2008tg} and  Tarrio and Vandoren \cite{Tarrio:2011de}, we included a dilatonic contribution to Critical Gravity, finding an asymptotically Lifshitz solution for $z=4$. Nevertheless, {this solution was very restrictive in $z$ and did not include the Schwarzchild-AdS with planar base manifold case as an isotropic limit}. A way to circumvent this obstacle was, in the spirit of \cite{Correa:2014ika,Ayon-Beato:2015jga,Ayon-Beato:2019kmz,Bravo-Gaete:2020ftn}, to add a non-minimally coupled scalar field $\Phi$ with a potential given by a mass term and a quartic term. Using this approach, we found a solution for $z$ defined in the interval $(1,4)$, recovering the Schwarzchild-AdS case with planar base manifold in the limit $z\rightarrow 1$. Moreover, when we analyzed the limiting case $z=4$, we found an additional solution that \textcolor{black}{represented, in fact, a stealth configuration overflying the $z=4$ solution found when we only considered the dilatonic contribution}. That is, this solution, while resulting in a vanishing $T_{\mu\nu}^{\Phi}$, includes a non-trivial field $\Phi$, a potential $U(\Phi) \sim \,  \Phi^4 $ and $\xi=1/6$}. As a result, both solutions exhibit different thermodynamical behaviors. {Indeed, we were able to show that, while both of them satisfy the First Law of Thermodynamics as well as the Smarr relation, {in this case the non-stealth configuration is thermodynamically preferred}. This was achieved by comparing their \textcolor{black} Helmholtz free energies}.\\
In this same context, one could wonder if these configurations also obey a higher dimensional Cardy formula \cite{BravoGaete:2017dso}. Indeed, by performing a double Wick rotation
$${t}\rightarrow i x_{1},\qquad {x}_{1}\rightarrow i t,$$
on a metric (\ref{eq:fbh})  and an adjustment of the location of the horizon in order to ensure the correct identification of its Euclidean version, one can calculate the solitonic masses using the quasilocal method described in \cite{Kim:2013zha,Gim:2014nba} and obtain that, for the $z=4$ solution (\ref{eq:sol_z4}),
$$
\mathcal{M}_{\mathrm{sol}}=-\frac {2 \sqrt{3} \,\Omega_2}{5 \kappa l},
$$
and that, for the solutions with a scalar field $\Phi$ (see equations  (\ref{eq:sol_z4}) and (\ref{solution})), the mass of the corresponding solitonic configuration, can be generally expressed as:
$$
\mathcal{M}_{\mathrm{sol}}=-{\frac {z\Omega_{{2}} \left( {l}^{2}\Lambda
 -{z}^{2}+4\,z\right) }{2 {l}^{3}\Lambda\,\kappa}}\,\left(\frac{2}{2+z}\right)^{\frac{z+2}{2}}.
$$
With all the above, we can verify that the generalized Cardy formula \cite{BravoGaete:2017dso}, which in $D$-dimensions reads
\begin{equation*}\label{generCardyd}
\mathcal{S}_\text{C}=\frac{2\pi l(z+D-2)}{D-2}
\!\left[\!\frac{-(D-2)\mathcal{M}_{\mathrm{sol}}}{z}\!\right]^{\frac{z}{z+D-2}}
\!{\mathcal{M}_{\mathrm{bh}}^{\frac{D-2}{z+D-2}}},
\end{equation*}
holds, that is $\mathcal{S}_{W}=\mathcal{S}_{\mathrm{C}}$.
\\
Further work would naturally include the extension of this analysis on different fronts, one of them, being the lifting to higher dimensions { \cite{Bra-Ju} where the conformal value in four dimensions $\xi=1/6$ \textcolor{black}{could be explained. Another} evident extension would be to analyze whether similar configurations can be obtained starting with a magnetic ansatz, as opposed \textcolor{black}{the} electric one. Additional future problems could \textcolor{black}{also} include the study of anisotropic configurations with conserved charges such as \textcolor{black}{electrically or magnetically} charged Lifshitz black holes, adding for example linear Maxwell fields \cite{BravoGaete:2019rci} or non-linear electrodynamics \cite{Plebanski:1968,Alvarez:2014pra,Stetsko:2020nxb,Stetsko:2020tjg}}. In addition, and following \cite{Edery:2019bsh}, we can explore other gravity theories with critical conditions via the Weyl tensor and the square of the Ricci scalar.  {Together with the above, and as was shown in \cite{Moreira:2021iwy} , we can to study  the existence of static soliton-like structures on the Lifshitz spacetime (\ref{eq:Lifshitz})}. Another idea would be to consider more general asymptotic geometries, for example
hyperscaling violation metrics which are, in general, described by \cite{Charmousis:2010zz}
\begin{align*}
ds^2_{H}=\left( \frac{l}{r} \right)^{\frac{2\theta}{D-2}}\left( -\frac{r^{2z}}{l^{2z}}dt^2+\frac{l^2}{r^2}dr^2+\frac{r^2}{l^2} d\vec{x}^{2}  \right),
\end{align*}
where $\theta$ is known as the hyperscaling violation exponent.
Another interesting open problem would be to see whether these solutions can be generalized by the introduction of more dilatonic fields. Finally, it would be desirable, for completeness, to analyze the rotating case and provide more examples to test the Cardy-like formula
proposed in \cite{BravoGaete:2017dso}
\begin{eqnarray*}
\mathcal{S}_\text{C}&=&\pi
\sqrt{\frac{d_{\mbox{\tiny{eff}}}+z}{z}}\Big[\left(-2\mathcal{M}_{\mathrm{sol}}\right)^z
\frac{1}{d_{\mbox{\tiny{eff}}}^{d_{\mbox{\tiny{eff}}}}}\Big]
^{\frac{1}{z+d_{\mbox{\tiny{eff}}}}}\nonumber\\
&\times& \left(\sqrt{\Theta}+(d_{\mbox{\tiny{eff}}}+z)
\mathcal{M}^{\mathrm{rot}}_{\mathrm{bh}}\right)^\frac{1}{2}\nonumber\\ &\times&\left(\sqrt{\Theta}
-(d_{\mbox{\tiny{eff}}}-z)\mathcal{M}^{\mathrm{rot}}_{\mathrm{bh}}\right)
^\frac{d_{\mbox{\tiny{eff}}}-z}{2(d_{\mbox{\tiny{eff}}}+z)}, \nonumber
\label{CardyHvm}
\end{eqnarray*}
where
$$\Theta=(d_{\mbox{\tiny{eff}}}+z)^2 \big(\mathcal{M}^{\mathrm{rot}}_{\mathrm{bh}}\big)^2-4d_{\mbox{\tiny{eff}}} z\big(\mathcal{J}^{\mathrm{rot}}_{\mathrm{bh}}\big)^2,$$
and $d_{\mbox{\tiny{eff}}}$ is the effective spatial dimensionality, which reflects the scale of the entropy $\mathcal{S}$ with respect to the temperature $T$ as
$$\mathcal{S}\propto T^{\frac{d_{\mbox{\tiny{eff}}}}{z}}.$$

\begin{acknowledgments}

{The authors would like to thank,  Eloy Ay\'on-Beato, Mokhtar Hassaine, Julio Oliva and Isaac Ram\'irez for useful discussion and comments on this work.} \textcolor{black}{The authors thank the
Referee for the commentaries and suggestions to improve
the paper.}
\end{acknowledgments}





\begin{thebibliography}{99}
\bibitem{Deser:1981wh}
  S.~Deser, R.~Jackiw and S.~Templeton,
  Annals Phys.\  {\bf 140}, 372 (1982)
  [Annals Phys.\  {\bf 281}, 409 (2000)]
  Erratum: [Annals Phys.\  {\bf 185}, 406 (1988)].
  doi:10.1006/aphy.2000.6013, 10.1016/0003-4916(82)90164-6


\bibitem{Deser:1982vy}
  S.~Deser, R.~Jackiw and S.~Templeton,
  Phys.\ Rev.\ Lett.\  {\bf 48}, 975 (1982).
  doi:10.1103/PhysRevLett.48.975

 \bibitem{Bergshoeff:2009hq}
  E.~A.~Bergshoeff, O.~Hohm and P.~K.~Townsend,
  Phys.\ Rev.\ Lett.\  {\bf 102}, 201301 (2009)
  [arXiv:0901.1766 [hep-th]].

\bibitem{Lu:2011zk}
  H.~Lu and C.~N.~Pope,
  Phys.\ Rev.\ Lett.\  {\bf 106}, 181302 (2011)
  doi:10.1103/PhysRevLett.106.181302
  [arXiv:1101.1971 [hep-th]].


\bibitem{Anastasiou:2017rjf}
G.~Anastasiou, R.~Olea and D.~Rivera-Betancour,
Phys. Lett. B \textbf{788} (2019), 302-307
doi:10.1016/j.physletb.2018.11.021
[arXiv:1707.00341 [hep-th]].


\bibitem{Anastasiou:2021tlv}
G.~Anastasiou, I.~J.~Araya, C.~Corral and R.~Olea,
JHEP \textbf{2021} (2021) no.7, 156
doi:10.1007/JHEP07(2021)156
[arXiv:2105.02924 [hep-th]].



\bibitem{Kachru:2008yh}
  S.~Kachru, X.~Liu and M.~Mulligan,
  Phys.\ Rev.\ D {\bf 78}, 106005 (2008)
  [arXiv:0808.1725 [hep-th]].




\bibitem{Pang:2009wa}
D.~W.~Pang,
JHEP \textbf{01}, 120 (2010)
doi:10.1007/JHEP01(2010)120
[arXiv:0912.2403 [hep-th]].

\bibitem{Ross:2009ar}
S.~F.~Ross and O.~Saremi,
JHEP \textbf{09}, 009 (2009)
doi:10.1088/1126-6708/2009/09/009
[arXiv:0907.1846 [hep-th]].

\bibitem{Kuang:2017rpx}
X.~M.~Kuang, E.~Papantonopoulos, J.~P.~Wu and Z.~Zhou,
Phys. Rev. D \textbf{97}, no.6, 066006 (2018)
doi:10.1103/PhysRevD.97.066006
[arXiv:1709.02976 [hep-th]].

\bibitem{Brito:2019ose}
F.~A.~Brito and F.~F.~Santos,
EPL \textbf{129}, no.5, 50003 (2020)
doi:10.1209/0295-5075/129/50003
[arXiv:1901.06770 [hep-th]].



\bibitem{Eling:2014saa}
C.~Eling and Y.~Oz,
JHEP \textbf{11}, 067 (2014)
doi:10.1007/JHEP11(2014)067
[arXiv:1408.0268 [hep-th]].





\bibitem{Lemos:2011gy}
J.~P.~S.~Lemos and D.~W.~Pang,
JHEP \textbf{06}, 122 (2011)
doi:10.1007/JHEP06(2011)122
[arXiv:1106.2291 [hep-th]].



\bibitem{Ge:2014eba}
X.~H.~Ge, H.~Q.~Leng, L.~Q.~Fang and G.~H.~Yang,
Adv. High Energy Phys. \textbf{2014}, 915312 (2014)
doi:10.1155/2014/915312
[arXiv:1408.4276 [hep-th]].

\bibitem{Maldacena:1997re}
  J.~M.~Maldacena,
  Adv.\ Theor.\ Math.\ Phys.\  {\bf 2}, 231 (1998)
  [hep-th/9711200].

\bibitem{AyonBeato:2010tm}
E.~Ayon-Beato, A.~Garbarz, G.~Giribet and M.~Hassaine,
JHEP \textbf{04}, 030 (2010)
doi:10.1007/JHEP04(2010)030
[arXiv:1001.2361 [hep-th]].


\bibitem{Ayon-Beato:2009rgu}
E.~Ayon-Beato, A.~Garbarz, G.~Giribet and M.~Hassaine,
Phys. Rev. D \textbf{80} (2009), 104029
doi:10.1103/PhysRevD.80.104029
[arXiv:0909.1347 [hep-th]].

\bibitem{Taylor:2008tg}
M.~Taylor,
[arXiv:0812.0530 [hep-th]].

\bibitem{Tarrio:2011de}
  J.~Tarrio and S.~Vandoren,
  JHEP {\bf 1109}, 017 (2011)
  doi:10.1007/JHEP09(2011)017
  [arXiv:1105.6335 [hep-th]].

\bibitem{Gibbons:1987ps}
G.~W.~Gibbons and K.~i.~Maeda,
Nucl. Phys. B \textbf{298}, 741-775 (1988)
doi:10.1016/0550-3213(88)90006-5

\bibitem{BravoGaete:2017dso}
M.~Bravo Gaete, L.~Guajardo and M.~Hassaine,
JHEP \textbf{04} (2017), 092
doi:10.1007/JHEP04(2017)092
[arXiv:1702.02416 [hep-th]].


\bibitem{KerrSchild}
R.~P.~Kerr and A.~Schild,
Proc. Symp. Appl. Math 17, 199 (1965).

\bibitem{Dereli:1986cm}
T.~Dereli and M.~Gurses,
Phys. Lett. B \textbf{171}, 209-211 (1986)
doi:10.1016/0370-2693(86)91533-9

\bibitem{Gibbons:2004uw}
G.~W.~Gibbons, H.~Lu, D.~N.~Page and C.~N.~Pope,
J. Geom. Phys. \textbf{53}, 49-73 (2005)
doi:10.1016/j.geomphys.2004.05.001
[arXiv:hep-th/0404008 [hep-th]].

\bibitem{Anabalon:2009kq}
A.~Anabalon, N.~Deruelle, Y.~Morisawa, J.~Oliva, M.~Sasaki, D.~Tempo and R.~Troncoso,
Class. Quant. Grav. \textbf{26}, 065002 (2009)
doi:10.1088/0264-9381/26/6/065002
[arXiv:0812.3194 [hep-th]].

\cite{Ett:2010by}
\bibitem{Ett:2010by}
B.~Ett and D.~Kastor,
Class. Quant. Grav. \textbf{27}, 185024 (2010)
doi:10.1088/0264-9381/27/18/185024
[arXiv:1002.4378 [hep-th]].

\bibitem{Ayon-Beato:2014wla}
E.~Ay\'on-Beato, M.~Hassa\"\i{}ne and M.~M.~Ju\'arez-Aubry,
Phys. Rev. D \textbf{90}, no.4, 044026 (2014)
doi:10.1103/PhysRevD.90.044026
[arXiv:1406.1588 [hep-th]].



\bibitem{Ayon-Beato:2010vyw}
E.~Ayon-Beato, A.~Garbarz, G.~Giribet and M.~Hassaine,
JHEP \textbf{04} (2010), 030
doi:10.1007/JHEP04(2010)030
[arXiv:1001.2361 [hep-th]].


\bibitem{Bravo-Gaete:2015xea}
M.~Bravo-Gaete and M.~Hassaine,
Phys. Rev. D \textbf{91} (2015) no.6, 064038
doi:10.1103/PhysRevD.91.064038
[arXiv:1501.03348 [hep-th]].




\bibitem{AyonBeato:2004ig}
E.~Ayon-Beato, C.~Martinez and J.~Zanelli,
Gen. Rel. Grav. \textbf{38}, 145-152 (2006)
doi:10.1007/s10714-005-0213-x
[arXiv:hep-th/0403228 [hep-th]].

\bibitem{AyonBeato:2005tu}
E.~Ayon-Beato, C.~Martinez, R.~Troncoso and J.~Zanelli,
Phys. Rev. D \textbf{71}, 104037 (2005)
doi:10.1103/PhysRevD.71.104037
[arXiv:hep-th/0505086 [hep-th]].


\bibitem{Ayon-Beato:2015qfa}
E.~Ay\'on-Beato, M.~Hassa\"\i{}ne and M.~M.~Ju\'arez-Aubry,
[arXiv:1506.03545 [gr-qc]].


\bibitem{Correa:2014ika}
  F.~Correa, M.~Hassaine and J.~Oliva,
  Phys.\ Rev.\ D {\bf 89}, no. 12, 124005 (2014)
  doi:10.1103/PhysRevD.89.124005
  [arXiv:1403.6479 [hep-th]].


\bibitem{Ayon-Beato:2015jga}
  E.~Ayón-Beato, M.~Bravo-Gaete, F.~Correa, M.~Hassaïne, M.~M.~Juárez-Aubry and J.~Oliva,
  Phys.\ Rev.\ D {\bf 91}, no. 6, 064006 (2015)
  doi:10.1103/PhysRevD.91.064006
  [arXiv:1501.01244 [gr-qc]].

\bibitem{Ayon-Beato:2019kmz}
  E.~Ayón-Beato, M.~Bravo-Gaete, F.~Correa, M.~Hassaine and M.~M.~Juárez-Aubry,
  arXiv:1904.09391 [hep-th].

\bibitem{Bravo-Gaete:2020ftn}
M.~Bravo-Gaete and M.~M.~Ju\'arez-Aubry,
Class. Quant. Grav. \textbf{37} (2020) no.7, 075016
doi:10.1088/1361-6382/ab7694
[arXiv:2002.10520 [hep-th]].





\bibitem{Wald:1993nt}
  R.~M.~Wald,
  Phys.\ Rev.\ D {\bf 48}, no. 8, R3427 (1993)
  doi:10.1103/PhysRevD.48.R3427
  [gr-qc/9307038].



\bibitem{Iyer:1994ys}
  V.~Iyer and R.~M.~Wald,
  Phys.\ Rev.\ D {\bf 50}, 846 (1994)
  doi:10.1103/PhysRevD.50.846
  [gr-qc/9403028].





 \bibitem{Kim:2013zha}
  W.~Kim, S.~Kulkarni and S.~-H.~Yi,
  Phys.\ Rev.\ Lett.\  {\bf 111}, no. 8, 081101 (2013)
  [arXiv:1306.2138 [hep-th]].

\bibitem{Gim:2014nba}
  Y.~Gim, W.~Kim and S.~-H.~Yi,
  arXiv:1403.4704 [hep-th].


\bibitem{Abbott:1981ff}
  L.~F.~Abbott and S.~Deser,
  Nucl.\ Phys.\ B {\bf 195}, 76 (1982).
  doi:10.1016/0550-3213(82)90049-9

\bibitem{Deser:2002rt}
  S.~Deser and B.~Tekin,
  Phys.\ Rev.\ Lett.\  {\bf 89}, 101101 (2002)
  doi:10.1103/PhysRevLett.89.101101
  [hep-th/0205318].

\bibitem{Deser:2002jk}
  S.~Deser and B.~Tekin,
  Phys.\ Rev.\ D {\bf 67}, 084009 (2003)
  doi:10.1103/PhysRevD.67.084009
  [hep-th/0212292].

\bibitem{Herrera-Aguilar:2020iti}
A.~Herrera-Aguilar, D.~F.~Higuita-Borja and J.~A.~M\'endez-Zavaleta,
[arXiv:2012.13412 [hep-th]].


\bibitem{Bravo-Gaete:2015iwa}
M.~Bravo-Gaete, S.~Gomez and M.~Hassaine,
Phys. Rev. D \textbf{92} (2015) no.12, 124002
doi:10.1103/PhysRevD.92.124002
[arXiv:1510.04084 [hep-th]].






\bibitem{Smarr:1972kt}
  L.~Smarr,
  Phys.\ Rev.\ Lett.\  {\bf 30}, 71 (1973)
  [Erratum-ibid.\  {\bf 30}, 521 (1973)].


\bibitem{Bra-Ju}
  M.~Bravo, M.~Ju\'arez and G.~Vel\'azquez, {\em{work in progress}}.




\bibitem{BravoGaete:2019rci}
M.~Bravo Gaete, S.~Gomez and M.~Hassaine,
Eur. Phys. J. C \textbf{79} (2019) no.3, 200
doi:10.1140/epjc/s10052-019-6723-6
[arXiv:1901.09612 [hep-th]].


\bibitem{Plebanski:1968}
  J.~Pleb\'anski,
  {\it Lectures on Non-Linear Electrodynamics} (Nordita, 1968).


\bibitem{Alvarez:2014pra}
A.~Alvarez, E.~Ay\'on-Beato, H.~A.~Gonz\'alez and M.~Hassa\"\i{}ne,
JHEP \textbf{06} (2014), 041
doi:10.1007/JHEP06(2014)041
[arXiv:1403.5985 [gr-qc]].


\bibitem{Stetsko:2020nxb}
M.~M.~Stetsko,
[arXiv:2012.14902 [hep-th]].


\bibitem{Stetsko:2020tjg}
M.~M.~Stetsko,
Gen. Rel. Grav. \textbf{53} (2021) no.1, 2
doi:10.1007/s10714-020-02777-w
[arXiv:2012.14915 [hep-th]].

\bibitem{Edery:2019bsh}
A.~Edery and Y.~Nakayama,
JHEP \textbf{11} (2019), 169
doi:10.1007/JHEP11(2019)169
[arXiv:1908.08778 [hep-th]].

\bibitem{Moreira:2021iwy}
D.~C.~Moreira,
[arXiv:2111.04487 [gr-qc]].


\bibitem{Charmousis:2010zz}
  C.~Charmousis, B.~Gouteraux, B.~S.~Kim, E.~Kiritsis and R.~Meyer,
  JHEP {\bf 1011}, 151 (2010).


\end{thebibliography}
\end{document}